# The offline digital currency puzzle solved by a local blockchain

HENRIQUE DE CARVALHO VIDEIRA[1]

henriquevideira@live.com | henrique.carvalho@coppead.ufrj.br

https://orcid.org/0000-0001-9362-2244

Federal University of Rio de Janeiro. Rua Pascoal Lemme,355. Ilha do Fundão, Rio de Janeiro–RJ

Central Bank of Brazil. Avenida Presidente Vargas,730. Centro, Rio de Janeiro-RJ.

## ABSTRACT

A major drawback in deploying central bank digital currencies (CDBC) is the offline puzzle, which requires that a CBDC must keep the provision given by cash, and, simultaneously, avoid double-spending, counterfeiting, and other issues. The puzzle is solved by minting the coins in serials, which are stored on a local blockchain (e.g. smartphone). The local blockchain is secured by keys embedded in the hardware and can be continuously mined by the wallet to enhance security. The coins can be either minted as hot coins, which can be retrieved in case of loss, or minted as cold coins, like physical cash.

## JEL CODES

E42, E51, E58, O33, O38

## KEYWORDS

offline CBDC, offline cash, hot coin, cold coin, digital non-repudiation

---

[1] Central Bank of Brazil, Auditor of the Brazilian Financial System, and a guest member of the Digital Brazilian Real Initiative. Federal University of Rio de Janeiro, PhD in Finance.
The views expressed in this article are those of the author and do not necessarily reflect the view of the Central Bank of Brazil.

1. **INTRODUCTION**

The offline feature of a Central Bank Digital Currency (CBDC) is one of the most challenging problems to solve in its design. Even though most transactions nowadays are processed online, communication with a third party may not be always available, inevitably requiring physical cash. Therefore, a CBDC must present the same provision given by cash while being offline, which is the greatest drawback of current CBDC projects. In an offline cash environment, major challenges that must be addressed are well known (Chu et al. (2022)), such as double spending, non-repudiation, unforgeability, and replay attacks.

Before discussing the offline cash approaches derived from industry and literature, it is worth giving a brief explanation between coins and tokens. Coins are either minted by an authority (i.e. a CBDC) or mined using confirmation protocols, such as Bitcoin (Nakamoto(2008)) and Ethereum (Buterin(2019)), which require nowadays a proof-of-work (PoW) and a proof-of-stake (PoS), respectively. Therefore, one can claim that a coin has its own proprietary blockchain. On the other hand, tokens are digital versions of physical and financial assets that can be claimed and exchanged in an existing blockchain. Furthermore, tokens usually need a medium of exchange to be settled (e.g., deposit tokens and wholesale tokens), but could also be exchanged as bearer instruments (Garrat and Shin (2023)). In the settlement, tokens have a face value that can be either exchanged at par, may entail slight variations, or might have an unconstrained fluctuation over a coin or stablecoin.

Aiming at a solution for common issues in offline cash, several initiatives have been discussed over the years. The first known initiative to deploy offline cash began with Mastercard and Visa in the 90s, but both attempts did not succeed. As seen in Bátiz-Lazo and Moretta (2016), the main reason for failure was the lack of universal acceptance by retailers and consumers, which kept some transactions still in cash rather than using e-cash. Therefore, as consumers still had to hold cash and e-cash simultaneously, their option was the universally accepted option, which is the physical cash. Besides the lack of universality, users did not perceive any other benefits derived from e-money, which inevitably led the project to fail. However, there is one interesting fact regarding the Mastercard initiative when tested in a university: inside the university, the e-money card was simultaneously the ID university card, the library card, and other conveniences. Thus, as the students perceived a wide benefit in using the card, acceptance was high in the university. Thus, one may infer that an offline design must present enhanced functionalities in order to make the users exchange the "king cash" with the e-cash.

With respect to CBDC, one of the first experiences was launched in Uruguay (Sarmiento (2022)). An offline e-peso project was designed by the Central Bank of Uruguay in a pilot project that lasted 6 months. In this project, transactions occurred using smartphones and each token issued had a serial number associated with a user. This serial number could derive other serials whenever a request for fractional payment occurred. The electronic notes could be used at a registered store or exchanged between users (P2P). The e-pesos were secured by a central authority whether users eventually lost their smartphones or wallets. Although many enhancements were deployed, it was not, in fact, an offline technology because the underlying infrastructure was fully dependent on a telecommunication network available full-time.

The Bank for International Settlement (BIS) has also presented some initiatives related to CBDCs, such as the project Helvetia (BIS (2020)) and the project Aurum (BIS (2022)). The Helvetia initiative uses a CBDC to settle payments in cross-tokens operations, replacing the typical settlement given by stablecoins. In the project Aurum, there is a two-tier infrastructure like the one adopted in this article, comprised of an interbank market (wholesale) and the retail market (e-wallet). However, Aurum has an interesting novelty: tokens in the retail market can be stablecoins supported by a CBDC, which can mitigate the risk of stablecoins supported by non-regulated assets.

Another leading initiative is the e-krona project launched by the Central Bank of Sweden in 2019 (Sveriges Riksbank (2019)). In one of its reports, the project lists the main concerns regarding offline cash: double-spending; lost offline wallet; the threshold for e-cash stored offline; consecutive transactions that should be enabled offline; and the maximum duration for a user to stay offline. However, these issues are not addressed yet by the project.

Before reaching the premises of the design, it is worth mentioning the common issues related to offline cash found in the literature, though some of them have already been introduced.

## 1.1. Offline Issues

### 1.1.1. Double spending

As digital tokens are merely an array of bits, this data in the offline wallet can be copied by malicious citizens and used indefinitely, whether a countermeasure is not properly addressed. In order to solve this issue, Eslami and Talebi (2011) propose an untraceable offline e-cash system using the ElGamal signature scheme. In this scheme, double spending can be prevented by revealing the identity information of malicious spenders to other users. However, this approach does not avoid double spending at its origin, enabling counterfeited transactions until being discovered through the ElGamal signature by the nodes. Another approach that does not avoid double spending at this origin is given by Dmitrienko et al. (2017), where malicious users are only identified after they double-spend the bitcoins. A viable solution for double spending is given by Christodorescu et al. (2020), which is specific to a CBDC design. This solution is supported by a trusted execution environment (TEE) with incremented counters for each transaction, both on the payer and the payee. This article is based on their solution as a starting point, enhancing new features yet to be discussed.

### 1.1.2. Non-repudiation

Non-repudiation, as defined by Kremer et. al (2002), is the undeniable will of having participated in part or the whole of a communication. Therefore, an efficient non-repudiation mechanism avoids a malicious receiver from denying a successful transaction. A non-repudiation mechanism is part of the current solution, which is better explained in further sections.

### 1.1.3. Verifiability

Verifiability is the feature of auditing offline transactions. This requirement is supported online by Bitcoin and Ethereum, as every confirmed transaction is recorded in blocks. In this article, the tracking history of the coin is stored in its unique chain.

### 1.1.4. Anonymity

Anonymity is often discussed as a feature that every digital cash should present. However, anonymity is the main driver for money laundering and illegal activities. On the other side, one can advocate that the "king cash" offers protection against illiquidity in the financial system - a feature named provision. In fact, even in Bitcoin and Ethereum, there is no full anonymity because public keys are published in every transaction. Thus, as Möser (2013) defined, a correct definition is pseudonymity. Pseudonymity presents a security flaw whether one keeps using the same public key in every transaction, turning more feasible for a hacker to make the link between the user and the public key. The solution for this issue can be provided by the node that maintains the account, changing randomly the public key, as suggested by Reid and Harrigan (2013). As the current solution is based on a unique chain for each coin holding several public keys, the randomization of public keys is recommended.

### 1.1.5. Unforgeability

Money cannot be minted in "thin air", but rather each CBDC must be issued by its respective Central Bank regarding uniqueness. The e-krona and the e-peso project mitigate this issue by assigning serial numbers to each digital coin, as discussed in Armelius et al. (2020). However, assigning serial numbers is not enough to ensure unforgeability in replay attacks. In replay attacks, the communication between the sender and the receiver can eventually be blocked, avoiding the transfer of the coin. As communication is blocked, the sender also does not receive the cancellation message, which can retrieve the coin back to him. A moment later, the frozen transaction is maliciously released to the receiver. Upon receiving the transaction, the transfer will not be rejected because signatures are still valid: therefore, a coin through "thin air" is forged, which can be replayed several times. The solution given by Christodorescu et al. (2020), which is also employed in this article, mitigates this issue by using counters that are automatically incremented in each transfer.

**1.2. Design of the Solution**

The article deals with the offline puzzle by authenticating a personal device (e.g. smartphone) in a secure environment, which is provided by keys embedded in the microchip. Each coin is serialized and stored on the personal device as a local blockchain. This chain can be either static or dynamic, in which the latter requires the wallet (device) to keep mining the coin continuously. Furthermore, the novel solution provides new enhancements, such as the "hot coin" (loss of coin), "cold coin" (store of value), and the non-repudiation mechanism. These features are explained in detail in the next section, which contains the Methodology. The remainder of the article presents the Conclusion, References, and, finally, an Appendix containing a typical block of the coin's chain.

**2. METHODOLOGY**

The current approach seeks a solution for offline cash, which can enable a legal tender CBDC, though other crypto coins like Bitcoin, Ethereum, or other stablecoins could also benefit from this offline methodology. Thus, there is a Central Bank issuing serialized coins (Proof-of-Authority), which turns automatically into its liability. Before presenting the solution, the following premises are crucial for acknowledging the design.

**2.1. Basic Premises**

2.1.1. The ledger is a two-tier permissioned DLT, like the one discussed in Araujo (2022), where a CBDC is delivered by an authorized Bank. Before delivering the CBDC to the citizen in exchange for his tokens, the Bank had previously made a fiat currency withdrawal from its reserve account in the Central Bank. The DLT containing the coin's chain can be either publicly available or maintained in a private ledger managed by the Central Bank and the Financial System (Banks). If the ledger is public, security must be strengthened by public keys that are changed on a regular basis, as discussed before. Regardless of being public or private, the ledger has a Proof-of-Authority (POA) validation mechanism, which can be subject to political and legal scrutiny.

2.1.2. Communication between parties in offline mode is done using NFC (near-field communication) or Bluetooth.

2.1.3. The wallets used in the transactions could be either EMV-PIN physical cards or smartphones. Transfers between users can happen only whether one of them has at least one smartphone. Furthermore, for full functionality regarding repudiation, the holder of the EMV card must also carry a smartphone for verifying whether the transaction or successful or not.

2.1.4. Moreover, these electronic wallets must have a pair of public-private keys embedded inside the microchip that makes each customer wallet unique, which is the genesis of the TEE (trusted execution environment) extensively used in Christodorescu et al. (2020). TEE is a robust solution because potential attackers would have to extract the hardware keys directly from the microchip using heavy methods such as ion beams, microchip decapsulation, and scanning microscopes. These procedures would be costly and ineffective, as extracting limited CBDC coins would not compensate for these invasive methods.

2.1.5. Identity and Signatures are supported by public-key encryption. In this approach, there is a pair of keys held by each user, which are the public and the private key. The private key is kept secret by the holder and is used to sign a document, which is verified by the correspondent public key of the asymmetric pair (see Lin (2009)). Both keys are mathematically related, and one can assume that is unfeasible to derive a private key knowing the public key. The asymmetric keys that are generated in the current approach are the following:

   a. The financial institution, bank, or authorized node holds a pair of asymmetric keys, whose public key is signed by the Central Bank. Central Bank possesses the root certificate that is the origin of the chain of trust that is branched down to the citizen, whose asymmetric keys are central.vk (verify-public) and central.sk (sign-private). Central Bank public keys might be changed on a regular basis, which requires that the offline wallet must go online at the same time-frequency to update certificates. However, a database with previous central.vk keys is stored for verifying previous serial numbers signed by the old private key of the respective pair. For practical purposes, the entity where the customer owns an account balance is designated by "Bank", whose public and private keys are denoted by bank.vk (verify) and bank.sk (sign), respectively.

   b. Going below the chain of trust, the next branch is the unique wallet belonging to each user, which could be an EVM credit card or smartphone. As discussed before, TEE is an environment supported by asymmetric keys embedded in the hardware's wallet, whose public key is signed by the manufacturer of the microchip. For practical purposes, the certification given to the secured application layer (TEE) running in the secured hardware is named "wallet". The wallet is authenticated after an OEM verification over the hardware keys embedded in the microchip and, simultaneously, over the application layer of the TEE (see Christodorescu et al. (2020) for more detail). Finally, there is a public key and a private key assigned to the wallet, which is the wallet.vk and wallet.sk, respectively. In a dynamic chain, yet to be discussed in detail, blocks are continuously signed by the wallet.sk without requiring the user's signature.

c. During the online validation session with the Bank, the citizen itself assigns a PIN code, password, or biometric data that is used to generate another pair of asymmetric keys: Alice.vk and Alice.sk, which are the public and the private keys of a hypothetical citizen named Alice who intends to send a coin (sender). With respect to the receiver, which could be a hypothetical citizen named Bob or a PoS (point-of-sale), the keys would be like the ones already addressed to Alice (the sender): Bob.vk and Bob.sk. Due to didactic reasons and practical purposes, the receiver is only Bob.

d. At last, the Bank emits a certificate for both the wallet and Alice (or Bob). The wallet's certificate must include Alice's personal public key, which links the unique wallet to its authorized user. The same approach is used for Alice's personal certificate, which must include the public key of the wallet. Even though Alice's certificate could not have the public key of the wallet, this practice is recommended for enhancing security and anonymity.

2.1.6. If not otherwise indicated, signatures provided by Bob or Alice are signed first by their wallets (wallet.sk) and, over this hash, their personal signatures are used to sign (Alice.sk or Bob.sk).

## 2.2. Online Payment Protocol

The following procedures describe the sequential steps for the online protocol between Alice, her Bank, and the Central bank. The online protocol is triggered to mint the offline coins as soon as Alice sends a coin requirement to her Bank.

*2.2.1. Authentication and coin requirement*

Minting the coin should be the first step, but before that, Alice must authenticate her wallet and herself in her Bank as discussed in basic premises. In this authentication procedure, a certificate is emitted for Alice and for her wallet, which are linked to each other. After authentication, Alice must send a coin requirement to her bank demanding to withdraw part of her deposited tokens in exchange for offline coins. For a practical example in this article, Alice will cash out 100 tokens from her account, which corresponds to 100 coins (e.g., US$ 100) to be deposited in her offline wallet.

*2.2.2. Mint*

After succeeding in the previous procedure, the coin is minted by Central Bank in this step. The Bank establishes a secure connection with Central Bank cashing out 100 coins from its reserve account maintained in Central Bank, which is the exact offline cash demanded by Alice. At this moment, a coin with a unique serial number is minted by Central Bank on behalf of the Bank, starting the coin's chain with the Genesis block. This procedure is summarized in Figure 1, whereas the genesis block is shown in Table 1. As just as physical cash, a Central Bank must make use of a threshold for each minted coin (e.g., US$ 100), aiming at mitigating forgery and double spending risks. Moreover, an offline coin must be preferably employed in small payments (Sarmiento (2022)), letting online connections with greater transactions.

| Mint of the Coin | |
| --- | --- |
| Central Bank Public Key | The public key of the Central Bank (central.vk) that is linked to the serial number of the coin. Public keys of the Central Bank must be regularly updated in the citizen's wallet. |
| Serial Number | Unique Serial Number of the coin minted by the Central Bank. |
| Value | Any positive floating number. Alice's example: 100 coins required by the Bank on behalf of Alice |
| Timestamp | Minting Date and Time |
| Bank's Certificate | Public Key of the Bank (bank.vk) signed by the Central Bank (central.sk) |
| Hot Coin Expiration Date | The coin can only be spent until the expiration date. This is an option selected by Alice when she sends a coin requirement |
| Hot Coin Claim Deadline | The claim deadline is the time limit for claiming a coin without bearing the risk of losing the value to previous owners before the preceding one. This time limit can be automatically set by Central Bank |
| Hash | Hash of the previous fields |
| Signature of the Central Bank | Signature of the Central Bank over the Hash using its central.sk, confirming that tokens have been minted in central bank coins (CBDC) |

Table1: Genesis block (first block) of the coin's chain

This coin signed by the Central Bank could be a "hot coin" or a "cold coin", which depends on whether exists an expiration date or not for spending the coin, respectively. The hot coin, which has a deadline for transactions, grants a great chance for citizens to retrieve coins in their lost offline wallets. The chance of not retrieving a coin only occurs whether the previous holder also loses his wallet. The other type of coin, the cold coin, is just like physical cash in a wallet: the chance of retrieving a wallet with offline digital coins is the same as holding a leather wallet with metal coins. The option of choosing between hot coins and cold coins is selected by Alice when she sends a coin requirement to her Bank. This subject is discussed in more detail in further sections.

Along with the hot and cold coin, another novelty arises: each coin is stored in a local blockchain (offline blockchain), which means that a block is linked to the hash of another one. The local blockchain provides a tracking history of the coin, storing previous pseudonyms (public keys) and transactions. Finally, after the Genesis block that mints the coin, the consecutive block is the one that transfers 100 offline coins to Alice.

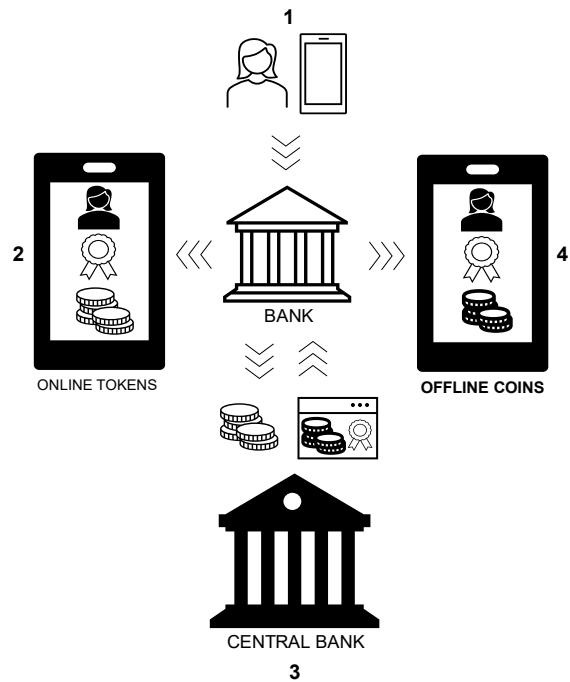

Figure 1: Minting offline coins requested by Alice.
A symbol comprised of 3 arrows is a message between two parties. The numbers labeled from 1 to 4 represent the sequence of events regarding the process. The black coin icon is related to offline coins, whereas the white one is a symbol for tokens.

### 2.2.3. *Offline coins delivered by the Bank.*

After the Genesis block, the Bank must transfer the 100 offline coins to Alice, keeping the online secure connection established before. In this step, the Bank will add a second block to the chain in order to transfer the coin to Alice's offline wallet. Therefore, the ownership of the coin is transferred to Alice when the Bank signs this second block using its private key. Besides the Genesis block, a typical consecutive block of the chain is shown in Table 2 (Appendix), which could be a continuously mined block by the current holder (see dynamic chain in section 2.5) or a transfer block. A transfer block, in the current example, could be either a transfer from the Bank to Alice (Bank -> Alice) or the transfer from Alice to Bob (Alice -> Bob).

## 2.3. Offline Payment Protocol

The following procedures describe how Alice engages in an offline transaction either with Bob or with a Point-of-Sale.

### 2.3.1. *Offline Invoice.*

Bob sends an invoice to Alice requesting 60 offline coins. In this invoice, Bob fulfills the fields of a typical block (Table 2 - Appendix) with respect to the transaction. In this step, Bob sends the invoice serial (invoice#) that ensures that the coin will not be accepted twice in his wallet, avoiding a replay attack with money out of thin air.

### 2.3.2. Sender evaluates the invoice and certificates.

Alice's wallet makes an analysis of Bob's certificates and whether the number of coins held in her wallet is enough for the transaction.

### 2.3.3. Fractioning Offline Coins.

Some of the coins might be fractioned in order to make an exact payment. In this case, the coin's chain is forked into two other chains containing the history of the mother chain until the split. Thus, the fork generates two different chains whose total value corresponds to the mother coin. A fork block in each of the derived two chains is identical to each other, except for the value regarding the child value (field in Table 2 – Appendix). Therefore, the first derived chain contains a fork block regarding 60 coins, while the second chain presents a fork block holding 40 coins. In these two fork blocks, the current holder remains Alice because this fork block is signed before the transfer block to Bob, due to the logic of the chain. This fractioning procedure is illustrated in Figure 2 along with other steps required to transfer the coin to Bob. Due to the complexity of claiming fractioned hot coins, fractioning shall be limited to cold coins.

### 2.3.4. Building the Transfer Block

After fractioning the coins, the next task is to build the transfer block containing the invoice sent by Bob. The transfer block contains a random number (secret.nonce#) automatically generated by Alice's wallet, which is kept in secret from Bob until he finally approves the full chain. This secret is hashed with the other fields, whose result remains unchanged regardless of whether revealing or not the secret to Bob. Thus, Alice signs over this immutable hash of the transfer block, which enables Bob to verify Alice's signature using Alice.vk. This procedure leaves a single mathematical operation to the last step, which is whether the hash function over all fields (including the secret nonce) is equal to the hash of the block itself - already sent and signed by Alice. Thus, the full chain with the transfer block, except for its secret.nonce#, is transferred to Bob. However, the transfer block can only be ultimately validated when the secret is revealed.

### 2.3.5. Verifying the Chain

After receiving the full coin, Bob's wallet makes an analysis over the full chain. Thus, Bob verifies every certificate and signature that correspond to each block, including the transfer block signed by Alice. As the transfer block contains a missing field, which is a random number (secret.nonce#), Bob's wallet compares whether Alice's signature over the hash of the transfer block is valid. As discussed before, evaluating whether the hash of the transfer block is equal to its respective field is done in the last step. Moreover, the

invoice# number in the transfer block is verified by comparing its value to the position in the protected memory of Bob's wallet. As discussed before, this procedure is part of a security measure that automatically increments a counter in the receiver's wallet, mitigating a replay attack.

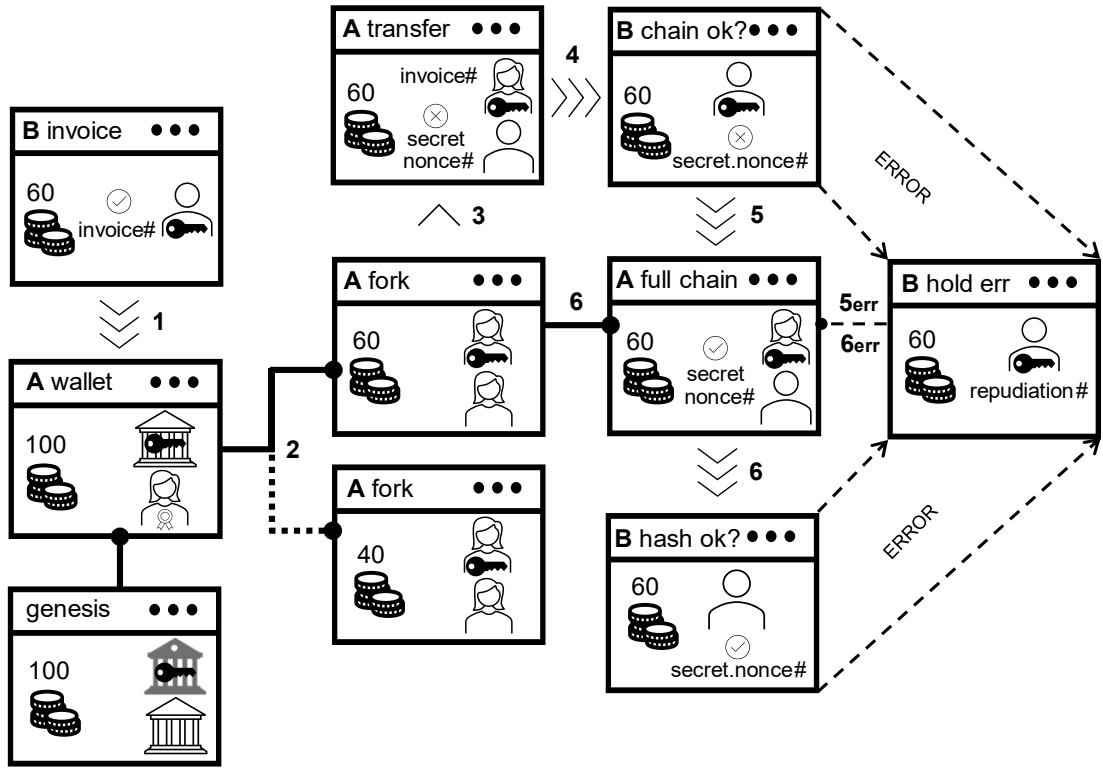

Figure 2: A full chain of a CBDC coin.
Letters "A" and "B" are abbreviations for Alice and Bob, respectively (e.g.: a block held by Alice is "A"). A symbol comprised of 3 arrows is a message between Alice and Bob, whereas a single arrow is an internal process inside Bob's or Alice's wallet. The numbers labeled from 1 to 6 represent the sequence of events regarding the main chain (continuous line) holding 60 coins. The dotted line represents the other branch of the fork that holds the remaining 40 coins of the wallet. A dashed line is a conditional event that is triggered whether an error occurs (repudiation#), returning 60 coins back to Alice. The secret.nonce# is kept in secret by Alice until Bob confirms that the full chain is ok, enabling the non-repudiation procedure.

When the analysis is complete, Bob's wallet requires him to insert his personal signature, which can validate or not the chain. Regardless of whether the chain is approved or not, an automatic procedure is triggered inside Bob's wallet when he inputs his personal signature: Bob signs over Alice's signature in the transfer block, keeping apart a repudiation signature (repudiation#) if any failure occurs (repudiation# is a field in Table 2 – Appendix). This failure can happen either when evaluating the full chain for the first time or when the secret.nonce# is revealed in a further step. Finally, if the coin's chain is approved, a signed confirmation message is sent to Alice; on failure, the repudiation# already signed is sent to Alice along with the signed error message.

If other coins are part of the payment, a successful transaction depends on the approval of every coin that makes part of the payment.

### 2.3.6. Receiver's (Bob) response

Even though last step has already presented some possible responses from Bob, it is worth discussing all possible scenarios:

a. No response comes from Bob's wallet after a certain period. In this case, as the secret.nonce# has not been revealed yet, the coin still belongs to Alice. The transfer block inside Alice's wallet is burned.

b. Bob sends a response to Alice's wallet, informing her that the chain is not approved. Thus, the rejection triggers a signed response to Alice including the repudiation# number. Bob's signature is verified by Alice's wallet, which burns the transfer block in its storage.

c. The coin's chain is approved. Even though full validation of the transfer block is not possible due to the missing secret.nonce#, Alice's signature over the hash of the transfer block is verified. At last, Bob sends a signed message confirming that every coin that is part of the payment is approved. If any of the coins that are part of the transaction presents a failure, it triggers the event shown in the previous topic.

### 2.3.7. Sender (Alice) sends the secret nonce.

After receiving the signed approval from Bob, Alice sends a signed message containing the secret.nonce# for each coin's chain. As soon as the secret.nonce# is delivered, Alice's wallet appends, permanently, the transfer block to the chain. At this moment, the coin does not belong to Alice anymore, remaining in her wallet until she goes online. When Alice goes online, the full chain containing the transfer block to Bob is uploaded to her Bank, which updates the ownership of the coin in the blockchain.

Even though the coin belongs to Bob in Alice's point of view, he can otherwise argue that an error has occurred in the transaction, driven by two events:

a. The secret.nonce# is incorrect because the hash function over all fields is not equal to the hash number that is shown in its proper field on the transfer block. However, this is a remote possibility, due to the ease in comparing the hash function over all fields to the hash already computed in the appropriate field.

b. The message has never been sent to Bob.

In both cases, a repudiation message (repudiation#) is already held on standby in Bob's wallet in order to return the coin back to Alice. Bob cannot avoid sending this repudiation, because this cancellation message was already signed by him in the previous stage. If Bob claims that something is happening with his connection to Alice's device, a QR code with the repudiation message must appear on his screen by his command. If Alice holds an EMV Card, she must take a photo with her smartphone over the QR code; after that, she can connect her EMV Card and her smartphone both to each other, transmitting the repudiation message. The worst scenario happens if Alice holds an EMV Card without a smartphone and, Bob, holds a smartphone. In this case, Alice will have to make a written copy of the string that corresponds to the QR Code. Other possible combinations, including the point-of-sale, are more favorable. Finally, as soon as Alice receives the repudiation message (repudiation#), which is the hash of the transfer block signed by Bob, the coin can be returned to her. The coin is ultimately released to Alice when the repudiation# is verified and written on the transfer block.

In a repudiation case, if Alice postpones loading repudiation# in her wallet due to a bad scenario, she must flag the coin with an error. This procedure will prevent these coins from being uploaded carrying Bob's ownership while Alice does not enter the repudiation message.

At last, Bob can claim that he has not received the coins, and neither a repudiation message is available. As explained above, these two events are mutually exclusive and cannot happen simultaneously.

## 2.4. Burning the Coins

The coin is burned into a token again depending on how some conditions are met and whether it is a hot coin or a cold coin. Moreover, a cold coin may be hold indefinitely, without being burned, whether the coin remains provisioned in cold storage.

### 2.4.1. Hot Coin

The great benefit of a hot coin is the great chance of retrieving the money in case of loss, which depends on the last holder not losing his wallet either. The hot coin is accepted by third parties until the expiration date defined in the mint block. After achieving the expiration date, the current holder must collect the coin until the claim deadline to not bear the risk of losing the coin. Regardless of the claim deadline, if the current holder has the full chain of the coin, it can be claimed immediately. The ownership of the full chain is confirmed whether some conditions are met:

a. If the current holder has not transferred the coin to another party, the transfer block will not appear in the coin's blockchain – the absence of a transfer block grants the ownership of the coin to the current holder.

b. If the current holder had made a failed attempt to transfer the coin to a third party, the repudiation signature will appear in the transfer block, assuring that the coin remains with the current holder.

After the claim deadline is met, the Central Bank verifies whether the coin was previously collected or not. If the coin or its remainders (fractioning) were not previously collected, the Central Bank will make an analysis of the uploaded chains. In this analysis, the longest valid chain of a coin is used to find the legitimate owner of the money.

### 2.4.2. Retrieving a lost Hot Coin

If the current holder has lost his wallet (smartphone or EMV card), he has a great chance of retrieving his lost coins, which can be automatically burned into tokens whenever possible. This mechanism is supported by the fact that every coin transaction remains in the sender's wallet even after the transfer is completed. Thus, when the last holder of the coin before the legitimate holder goes online, he will upload the chain containing the transfer block, which is the last block of the chain. If this transfer block belongs to the longest valid chain of the coin, Central Bank links the coin to the public key of the legitimate owner, retrieving the coin to him after the claim deadline.

Although retrieving lost coins is an enhancement compared to a fiat currency, a major issue is whether fractioning should be allowed or not in hot coins. This issue is derived from the complexity inherent to long fractioned chains, which can lead to excessive computing power in solving legitimate ownership. Thus, it seems reasonable that a citizen should carry hot coins for higher values (in case of losing the wallet) and cold coins for making fractional payments.

### 2.4.3. Cold Coin

Cold coins are designed to be identical to physical currency and do not have an expiration date or a claim deadline. If the current holder intends to burn the cold coin into tokens, he must prove that he is the legitimate owner of the public key shown as the beneficiary in the transfer block. On the opposite, if this coin was transferred before, this holder cannot claim this coin because there is a transfer block in the chain listing another beneficiary as the legitimate owner. This procedure is identical to collecting a hot coin before its claim deadline.

Even though the cold coin is the digital version of cash, there is one enhancement provided by the current approach, which is traceability. Unlike fiat currency, a cold coin provides a tracking history to the Central Bank whenever a citizen goes online. Although this tracking history can provide a powerful tool for facing money laundering and terrorism financing, the Central Bank cannot burn a dirty coin into the air due to these illegal activities. The main reason for not burning a coin into the air is that a third party with good faith may have received this coin.

Finally, the current design keeps "king cash" alive, as it enables the most important features of a fiat currency: its store of value, non-repudiation (legal tender), and liquidity, regardless of the owner. Moreover, people can remain safe against a dictatorial authority that could possibly seize their money, which is stored in cold coins protected by their personal signature.

## 2.5. Strengthening security with a dynamic chain

A possible measure for strengthening security is requiring that each coin's chain must be continuously mined, which is named dynamic chain. This mining process occurs inside each wallet using the Proof-of-Work (PoW) mechanism, which is analogous to Bitcoin.

This procedure requires that each wallet solves an arithmetical puzzle, which is finding a nonce number (mined.nonce#) that makes the hash of the block contain, at least, a certain number of zeros. The level of difficulty increases exponentially as more zeros are required to be found in the hash of the block (like Bitcoin's nonce). The level of difficulty must be calibrated by evaluating the coin's face value and the hardware where the coin is stored. The calibration can lead, for instance, to a chain that must produce new blocks every 10 seconds. At every 10 seconds, a newly mined block is appended to the chain signed only by the wallet's signature (wallet.sk).

When the coin is to be transferred to another party, the last block of the chain, which is the transfer block, is the one to be mined. Unlike a regular block, the transfer block contains the secret.nonce#, which is kept in secret from the receiver until he confirms the validity of the full chain. Therefore, the mined.nonce# in a transfer block is mined using the value of the secret.nonce# in addition to the other regular fields.

An issue occurs whether the wallet runs out of energy because mining inevitably stops. The calibration of difficulty (the number of zeros) can find the right balance between the cost of mining (hardware) and the face value of the coin. This cost barrier for counterfeiting can enable the computation of a variable mining speed. Therefore, if a wallet runs out of energy, the blocks that were not mined due to the blackout could be appended to the chain using an increased pace. The variable mining speed could be eventually hard coded, using the isolated hardware execution environment (TEE). Besides the blackout, another issue in continuity occurs whether the receiver sends a repudiation message canceling the transfer. In this case, there is a great chance that the chain loses its regular pace, requiring extra power allocation. Finally, a break in continuity might be so extensive that even an increased pace cannot solve the blocks that should have been mined. In this case, mining stops completely, which leads to blocking further transfers to other parties. As mining was interrupted, the remaining option is to collect the slow coin as a static coin, exchanging its value for tokens.

## 3. CONCLUSION

A central bank digital coin can only be regarded as a legitimate fiat currency whether it keeps its main features regardless of being online or offline. However, minting an offline digital currency without having double-spending, non-repudiation, and other issues is a puzzle. The current approach solves this puzzle by employing some strategies already found in the literature along with novel solutions.

The starting point of the article relies on the solution given by Christodorescu et.al (2020), which uses personal devices strengthened by hardware keys embedded in the microchip (trusted execution environment). This approach is part of the e-krona project, but the article leverages this solution with new enhancements.

The first enhancement provides a serial number for each coin, which is also a solution adopted by some CBDCs and some issued private coins. If the issued coin needs to be fractioned, its children carry the chain's history of the mother coin.

The second enhancement is the proper design of the transfer with respect to non-repudiation. The non-repudiation task starts with generating a random secret number (secret.nonce#), which is held in secret from the receiver until he approves the chain. Before even approving the chain, the receiver signs the hash of the transfer block, which is the repudiation signature (#repudiation). The repudiation# signature is automatically sent by the receiver's wallet whether any error occurs. This procedure ensures that there is always a repudiation# already signed before granting the coin, which unequivocally occurs when the secret.nonce# is revealed to the receiver.

The third enhancement is the most important one, which gives each coin a chain of blocks linked to each other through the hash of the previous one. This local blockchain contains the history of the coin, whose ownership tracking enables minting "hot coins" and "cold coins". A "hot coin" is a coin that is minted with expiration and claim deadlines. If a wallet is lost containing a hot coin, the legitimate owner can retrieve this coin if the previous owner does not lose his wallet either. This is an enhancement over cash, which is a physical bearer instrument that cannot be lost. On the other hand, the cold coin is much like cash, with exception to providing tracking history that is absent in cash. The tracking history in hot and cold coins provide a powerful tool for facing illegal activities, such as money laundering and terrorism financing. However, authorities cannot seize a coin's serial number, due to illegal activities, without bearing the risk of affecting a third party with good faith who received that coin. Another benefit of tracking history is the gathering of economic data, which can support proper public policies and research.

The third enhancement can be strengthened by a dynamic chain. A coin's chain becomes dynamic taking into account that each wallet is required to solve continuously an arithmetical puzzle. This puzzle is a Proof-of-Work challenge, like the Bitcoin's nonce, which requires that a mined.nonce# must lead to a hash containing, at least, a certain number of zeros. The required number of zeros depends on the face value of the coin and the wallet's hardware containing the coin. Therefore, each coin's chain must be continuously mined by the wallet of the holder, resulting in a long chain that is a proof of holding ownership. This dynamic chain carries a chunk of energy that can mitigate the financial benefit derived from possible counterfeiting.

The choice whether deploying a CBDC or not may be not an option anymore, due to the imminent tokenisation of everything (Heines et al. (2021)). In tokenisation, physical and financial assets are turned into digital versions of themselves, which can be claimed and settled in a blockchain. This new paradigm is a promise of more liquidity and transparency, whose enhanced trust is leveraged by smart contracts, mitigating part of the counterpart risk. However, tokenisation has a great issue, which is the lack of a cross-token settlement currency. In practice, the settlement has been made by stablecoins, which can be a source of financial instability (BIS (2019;2023)). Therefore, in order to mitigate this risk, and, simultaneously, provide new enhancements, the issuance of a smart CBDC has become even more crucial.

# 5. APPENDIX

| Typical Block | |
|---|---|
| Previous Signed Hash | Signed Hash of the previous block of the chain. |
| Timestamp | Date and time of the block |
| Current Holder's Certificate | The public key of the current holder of the coin signed by the Bank (Bank.sk)<br>Second Block (Bank -> Alice )    : Alice.vk (A ned by Bank.sk<br>Transfer Block (Alice -> Bob)    : Bob.vk signed by Bank.sk |
| Wallet's Certificate | The public key of the Current Holder's wallet (wallet.vk) signed by the Bank.<br>The offline wallet can be a smartphone or EMV card. |
| Bank's certificate | Public key of the Bank (bank.vk) signed by the Central Bank (central.sk) |
| Child Value (Fractioning) | The field remains empty until fractioning is required.<br>If this field presents a value, a fork has happened in this current block.<br>A fork occurs when a mother coin is fractioned into two required values, generating two children whose total sum corresponds to the mother coin.<br>The other child value is located on another chain, derived at the same fork. Both children keep their mother's chain history until the fractioning.<br>Fractioning is not recommended for hot coins due to claiming's complexity. |
| Invoice Serial invoice# | Transfer Block (Alice -> Bob )<br>Bob's invoice to Alice requesting an offline payment is numbered by using the unique counter number registered in the RPMB (Replay Protected Memory Block). This counter number is named invoice serial (invoice#), which is incremented at every written transaction in the protected memory of the TEE.<br>The invoice serial (invoice#) avoids a replay attack aiming at double-spending. |
| Mined Nonce mined.nonce# | This field is required for dynamic chains that require never-ending mining.<br>Mining happens continuously inside the wallet, which automatically solves the mined.nonce#.<br>In this local mining, mined.nonce# is the proof of solving an arithmetical puzzle by the current holder's wallet.<br>The hash of the mined block is signed only by the wallet (wallet.sk). |
| Secret Nonce secret.nonce# | This field only has a valid value in a Transfer Block (null in every other case).<br>In a transfer, the secret.nonce# is a random number generated by the current holder's wallet, which is kept in secret from the receiver until the coin is finally delivered. Even though the secret.nonce# is not provided to Bob at first, the hash of the transfer block that contains the random number in the hash function is given to Bob with the full chain. When Bob receives the full chain from Alice, an automatic procedure triggered by his wallet generates a cancellation message (repudiation#) that is put on hold. If any failure occurs, this cancellation message (repudiation#), which is Bob's signature (Bob.sk) over Alice's signature (Alice.sk), is sent back to Alice.<br>This procedure addresses the non-repudiation issue. |
| Hash | A Hash function over all the previous fields of the block. |

| Typical Block Signatures | |
|---|---|
| Previous Holder's signature | Second Block (Bank -> Alice )<br>The Bank uses its private key (Bank.sk) to sign over the Hash<br><br>Transfer Block (Alice -> Bob)<br>Alice's wallet signs over the Hash of the transfer block (wallet.sk). After that, this signed hash is signed over again using Alice's private key (Alice.sk), which is her PIN, password, or biometric data. This is the signature that transfers the coin's ownership to Bob. |
| Current Holder Cancellation Signature (repudiation#) | Transfer Block (Alice -> Bob)<br>This field is fulfilled whether a failure occurs.<br>As discussed in secret.nonce#, Bob's signature (#repudiation) over Alice's signature is always computed, regardless of whether he approves or not the full chain of the coin. If any failure occurs, the coin is released back to Alice through Bob's repudiation signature (repudiation#). |

Table 2. Typical Block of the Coin's chain.
A block can be appended continuously (dynamic) or only appended due to a transaction (static).